\documentclass[conference, twocolumn]{IEEEtran}

\usepackage{graphicx, subcaption} 
\usepackage{amsmath, amsbsy, amssymb, mathtools} 
\usepackage{tikz, pgfplots, circuitikz} 
\pgfplotsset{compat=newest} 
\usepgfplotslibrary{groupplots} 

\usepackage{hyperref} 
\usepackage{xcolor} 

\usepackage{epigraph} 
\usepackage[ruled, boxed, lined, linesnumbered]{algorithm2e} 
\usepackage{algpseudocode} 

\usepackage[group-separator={,}]{siunitx} 
\usepackage{float, dblfloatfix} 
\usepackage{mwe, booktabs, multirow, tabularx} 
\usepackage{physics, verbatim, wrapfig} 

\usepackage{pstricks} 

\hypersetup{
    colorlinks=true,
    linkcolor=blue,
    filecolor=magenta,      
    urlcolor=cyan,
}

\newcommand{\eg}{{\sl e.g., }}

\newcommand{\keywords}[1]{\textbf{Keywords:} #1}
\newcommand{\BfPara}[1]{{\noindent\bf#1.}\xspace}

\definecolor{color1}{RGB}{228,26,28}
\definecolor{color2}{RGB}{55,126,184}
\definecolor{color3}{RGB}{77,175,74}
\definecolor{color4}{RGB}{152,78,163}
\definecolor{color5}{RGB}{255,127,0}
\definecolor{color6}{RGB}{200,200,200}

\begin{document}

\title{\fontsize{16}{28}\selectfont Generative vs. Predictive Models in Massive MIMO Channel Prediction}
\author{
\large \textit{Ju-Hyung Lee} \ \ \large \textit{Joohan Lee} \ \ \large \textit{Andreas F. Molisch}
\\

University of Southern California 
} %
\maketitle

\begin{abstract}
Massive MIMO (mMIMO) systems are essential for 5G/6G networks to meet high throughput and reliability demands, with machine learning (ML)-based techniques, particularly autoencoders (AEs), showing promise for practical deployment. However, standard AEs struggle under noisy channel conditions, limiting their effectiveness. This work introduces a \emph{Vector Quantization}-based generative AE model (VQ-VAE) for robust mMIMO cross-antenna channel prediction. We compare \emph{Generative} and \emph{Predictive} AE-based models, demonstrating that Generative models outperform Predictive ones, especially in noisy environments. The proposed VQ-VAE achieves up to $15$ [dB] NMSE gains over standard AEs and about $9$ [dB] over VAEs. Additionally, we present a complexity analysis of AE-based models alongside a diffusion model, highlighting the trade-off between accuracy and computational efficiency.
\end{abstract}


\section{Introduction} 
\label{sec:intro}
The burgeoning demand for higher data rates and  reliable wireless communication has driven the adoption of massive MIMO (mMIMO) systems in Beyond-5G/6G networks. mMIMO systems leverages a large number of antennas at both the transmitter (TX) and receiver (RX) ends, enhancing spectral efficiency and system capacity. However, its performance depends heavily on accurate and efficient channel estimation. Traditional channel estimation relies heavily on explicit channel feedback using pilot signals (\eg DMRS, SRS, CSI-RS), which leads to substantial computational expense, particularly as the system scale increases.

In response to these challenges, both industry (\eg 3GPP Release 18 and 19) and academia have explored AI/ML-based methods for channel prediction and efficient feedback \cite{fire, 8640815, 9048929}. Among these, \emph{Autoencoder} (AE)-based approaches have gained popularity due to their simplicity, practicality, and solid performance.

Despite the advancements, existing AI/ML models, including \emph{Predictive} AE-based solutions, struggle in noisy environments where imperfect channel feedback degrades reconstruction accuracy at the RX. This paper delves into these challenges by proposing \emph{Generative} AE-based models (\eg , in particularly, to handle noisy conditions better in mMIMO communication systems.

\BfPara{Contributions}
Our key contributions are as follows:
\begin{itemize}
    \item \textbf{Generative vs. Predictive Models for mMIMO Channel Prediction:}  
    We compare the performance of AE-based generative and predictive models for mMIMO cross-antenna channel prediction. The results show that generative AE models outperform predictive AE models, especially under noisy conditions (see Fig.~\ref{fig:comparison}).

    \item \textbf{VQ-VAE for Robust Noisy Channel Prediction:}  
    We propose a Vector Quantization-based Generative AE model (VQ-VAE) \cite{vqvae} for mMIMO cross-antenna channel prediction. VQ-VAE achieves up to $\sim 15$ [dB] NMSE improvement over AE and $\sim 9$ [dB] over VAE in noisy conditions (see Fig.~\ref{fig:prediction}), while offering significantly lower computational complexity than the Diffusion model (see Table~\ref{table:complexity}).
\end{itemize}
\section{Background} \label{sec:Background}

\begin{figure*}[!ht]
    \centering
    \subfloat[Autoencoder]{\vspace{2.em} 
 \includegraphics[width=.75\columnwidth]{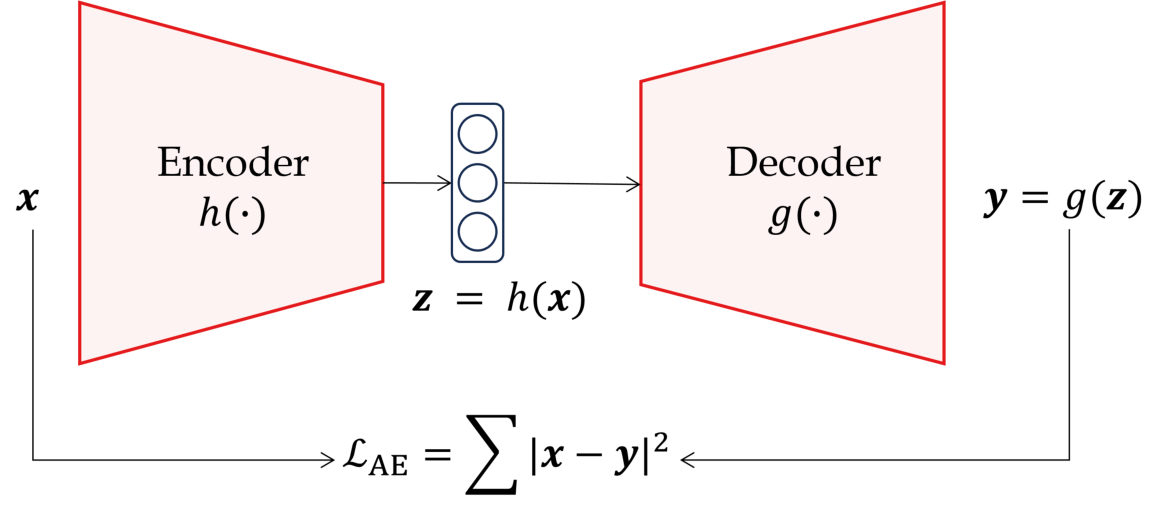}}
    \subfloat[Variational Autoencoder]{\vspace{0.em} \includegraphics[width=.95\columnwidth]{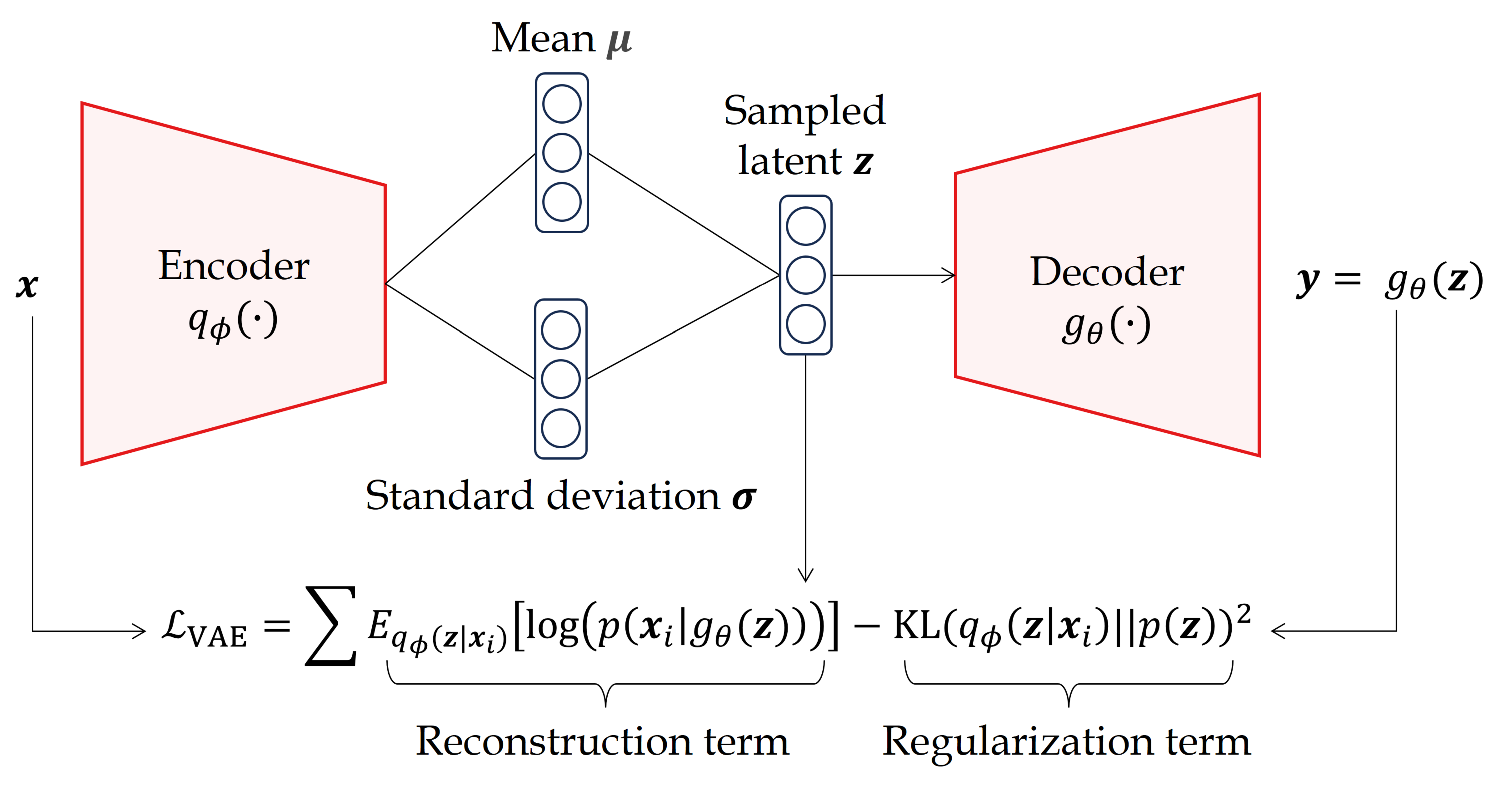}}
    \caption{AE vs. VAE model.}
    \label{fig:background}
\end{figure*}

\BfPara{Generative vs. Predictive Models}  
Generative models aim to estimate the joint probability distribution \(P(X, Y)\) of the input features \(X\) and the target variable \(Y\). From the joint distribution, marginal probabilities can be derived by summing over the other variable:  
\begin{equation}
    P(X) = \sum_{y} P(X, Y = y), \quad P(Y) = \sum_{x} P(X = x, Y).
\end{equation}
With these distributions, the conditional probability \(P(Y  |  X)\) can be computed using Bayes' theorem:
\begin{equation}
    P(Y  |  X) = \frac{P(X  |  Y) \cdot P(Y)}{P(X)}.
\end{equation}
This allows generative models not only to predict the likelihood of outcomes \(Y\) given inputs \(X\), but also to generate new data samples by capturing the underlying structure of the joint distribution.

In contrast, predictive (or discriminative) models directly estimate the conditional probability \(P(Y  |  X)\) without attempting to model the joint distribution \(P(X, Y)\). These models focus solely on predicting the target \(Y\) given an observation \(X\), prioritizing accurate predictions over a comprehensive understanding of the data distribution.

\BfPara{AE vs. VAE}  
Autoencoder (AE) \cite{ae} and Variational Autoencoder (VAE) \cite{vae} differ fundamentally in how they handle latent representations and data generation. 

\begin{enumerate}
    \item \textbf{Autoencoder (AE):}  
    An AE consists of an encoder and a decoder network. The encoder maps input data \(x\) into a low-dimensional latent vector \(z\) using an encoding function \(h(\cdot)\):
    \[
        \vb*{z} = h(\vb*{x}).
    \]
    The decoder reconstructs the input from the latent vector using a decoding function \(g(\cdot)\):
    \[
        \vb*{\hat{x}} = g(\vb*{z}) = g(h(\vb*{x})).
    \]
    The latent vector \(z\) serves as a compressed, deterministic representation of the input. AEs minimize the reconstruction error, typically using mean squared error (MSE) or cross-entropy loss:
    \[
        \mathcal{L}_{\mathrm{AE}} = \sum_{x \in D} \mathcal{L}(x, g(h(x))),
    \]
    where \( \mathcal{L}(x, y) = |x - y|^2 \) or cross-entropy. AEs excel at dimensionality reduction and reconstruction but lack the ability to generate diverse outputs beyond reproducing inputs.

    \item \textbf{Variational Autoencoder (VAE):}  
    A VAE introduces stochasticity by learning a probabilistic distribution over the latent space, enabling both reconstruction and data generation. Instead of encoding inputs to a fixed vector, the encoder outputs the parameters of a Gaussian distribution:
    \[
        h_{\phi}(x) = \left[\mu(\vb*{h}_{\text{enc}}), \log\sigma^2(\vb*{h}_{\text{enc}})\right],
    \]
    where \( \mu \) and \( \sigma^2 \) represent the mean and variance, respectively. The latent vector \(z\) is sampled using the reparameterization trick to maintain differentiability:
    \[
        \vb*{z} = \mu(\vb*{h}_{\text{enc}}) + \sigma(\vb*{h}_{\text{enc}}) \odot \epsilon, \quad \epsilon \sim \mathcal{N}(0, I).
    \]
    The decoder reconstructs the input from the sampled \(z\):
    \[
        \vb*{\hat{x}} = g_{\theta}(\vb*{z}),
    \]
    where \(g_{\theta}(\cdot)\) is the decoder network. VAEs optimize a loss function that balances reconstruction accuracy and latent space regularization:
    \begin{align}
          \mathcal{L}_{\mathrm{VAE}} =& \sum_{i} E_{q_{\phi}(z|x_i)}\left[ \log(p(x_i|z)) \right] \nonumber\\
        &- \mathrm{KL}(q_{\phi}(z|x_i) ||p(z)) 
    \end{align}
    The Kullback-Leibler (KL) divergence term ensures that the learned latent distribution \(q_{\phi}(z  |  x)\) remains close to the prior \(p(z)\), typically a standard normal distribution.

\end{enumerate}

AEs focus on deterministic reconstruction by learning the conditional distribution $p(y | x)$, limiting their ability to generate new data. In contrast, VAEs incorporate the joint distribution \(p(x, z)\) and sample from the latent space, allowing them to generate novel outputs that share patterns with the training data. 

\BfPara{Limitations in mMIMO Channel Prediction}  
Efficient channel prediction is essential in mMIMO systems to reduce computational overhead and ensure practical deployment. While ML-based approaches offer promising solutions, they do not address all challenges. Channel information (or reference signals) can often be noisy due to factors such as interference, hardware impairments, and estimation errors. As a result, it becomes crucial to develop noise-robust models that can maintain prediction accuracy while balancing computational complexity—our primary focus in this work.

Generative models, unlike predictive models, leverage their ability to learn underlying patterns in the data, enabling them to make accurate predictions even with imperfect or noisy inputs. This pattern-based understanding allows generative models to go beyond the raw input features, making them particularly well-suited for handling the uncertainties present in noisy wireless communication environments.

\section{VQ-VAE for mMIMO Channel Prediction} \label{sec:Body}

\begin{figure*}[!ht]
    \centering
    \includegraphics[width=1.3\columnwidth]{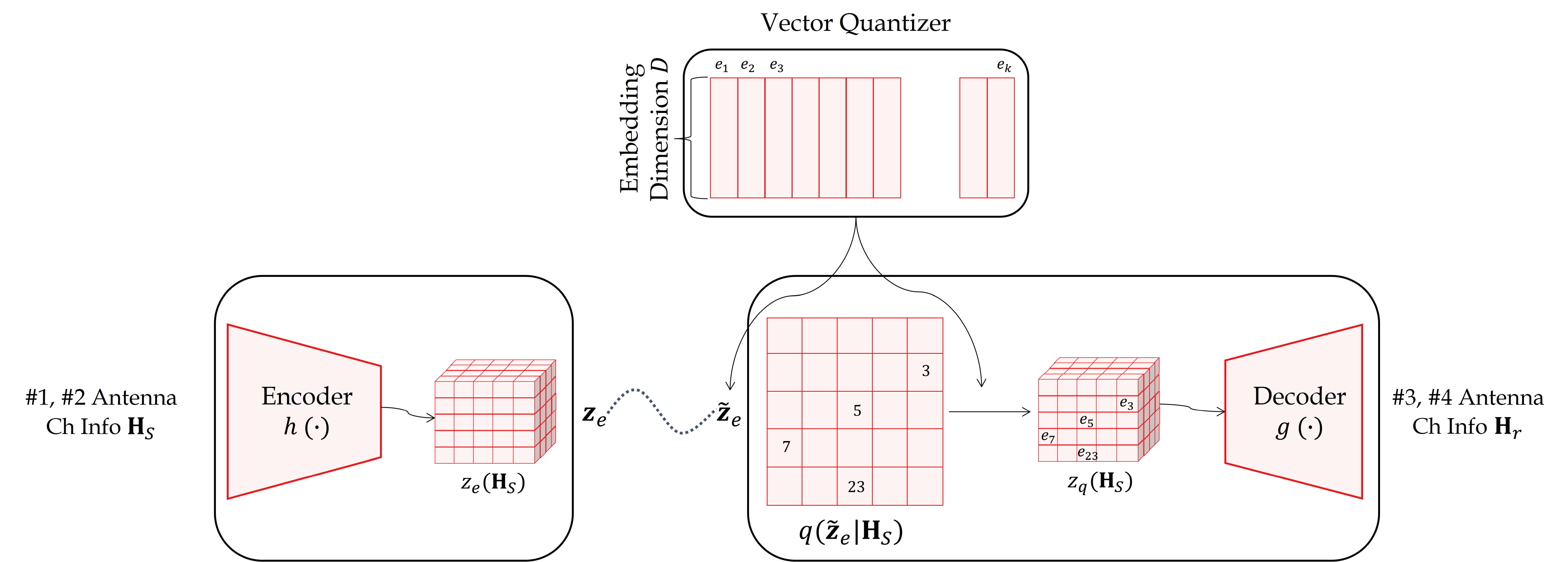}
    \caption{VQ-VAE based mMIMO channel prediction.}
    \label{fig:overview}
\end{figure*}

\subsection{Task: Cross-Antenna Channel Prediction}
Accurate estimation of the wireless channel, denoted as $\mathbf{H} \in \mathbb{C}^{M \times N}$, is essential for reliable communication in wireless systems, where $M$ and $N$ represent the number of transmitting and receiving antennas, respectively. 
Imperfect channel estimation introduces errors that degrade the signal quality, resulting in lower data rates and compromised reliability. 

In the uplink (UL) scenario, each user equipment (UE) transmits $M$ pilot symbols—one for each antenna—and the base station (BS) measures the received signals to estimate the corresponding channel coefficients. Using Sounding Reference Signals (SRS) or similar mechanisms, this process produces channel estimates $\mathbf{H}_{k} \in \mathbb{C}^{M \times 1}$ for each antenna $k$, where $k \in \{1, \ldots, K\}$ represents the index of the BS antennas. The total feedback for all antennas scales as $\mathcal{O}(M \times K)$, resulting in considerable overhead for large $M$ and/or $K$. As the number of antennas increases, this overhead becomes a major bottleneck, particularly in massive MIMO (mMIMO) systems with millimeter-wave (mmWave) frequency bands, where efficient communication is critical.

This work proposes a \emph{Cross-Antenna} channel prediction strategy to alleviate the overhead especially in RX. Instead of estimating channels for all antennas directly, we employ a neural network (NN)-based approach to predict the channel responses of some antennas using the channel information from a subset of others. More precisely, given channel measurements from a subset of $M_s$ antennas, denoted as $\mathbf{H}_s \in \mathbb{C}^{M_s \times 1}$, the neural network predicts the channel coefficients for the remaining $M_r$ antennas, $\mathbf{H}_r \in \mathbb{C}^{M_r \times 1}$. 

In our experimental setup, we train the model to predict the channel responses for a group of $M_r = 2$ antennas, using only the observed information from another group of $M_s = 2$ antennas. This demonstrates the feasibility of exploiting partial channel information to reduce the overall pilot overhead without compromising the quality of channel estimation.

\subsection{Approach: Vector Quantized VAE (VQ-VAE)}
As illustrated in Figure~\ref{fig:overview}, the encoder compresses the channel information from a subset of antennas, denoted as \(\mathbf{H}_s\), into a latent vector \(z_e\) using a series of convolutional layers, reducing the dimensionality of the input. This latent vector \(z_e\) is transmitted to the decoder, but may be affected by noise during transmission, resulting in a noisy version \(\tilde{z}_e\). 

The vector quantizer maps \(\tilde{z}_e\) to the nearest embedding from a predefined codebook, denoted as \(e_i\), where \(1 \leq i \leq k\) and \(k\) is the total number of codebook entries. The quantized vector \(z_q\) is then passed to the decoder, which reconstructs the original channel information from \(\mathbf{H}_s\). 

By transmitting latent vectors instead of raw channel information, the proposed encoding-decoding process reduces the overhead and improves robustness to noise—further discussed in the next section.

\subsection{Evaluation \& Training}
The reconstruction accuracy is evaluated using the normalized mean squared error (NMSE), defined as:
\begin{align}
    \mathrm{NMSE}(\vb*{x}, \hat{\vb*{x}}) &= \frac{\mathrm{MSE}(\vb*{x}, \hat{\vb*{x}})}{\mathrm{Var}(\vb*{x})}, \\
    \mathrm{MSE}(\vb*{x}, \hat{\vb*{x}}) &= \frac{1}{N} \sum_{i=1}^N (x_i - \hat{x}_i)^2,
\end{align}
where \(\vb*{x}\) is the ground-truth data, \(\hat{\vb*{x}}\) is the predicted data, and \(N\) is the number of samples. The variance \(\mathrm{Var}(\vb*{x})\) normalizes the MSE, making NMSE invariant to data scaling.

AE, VAE, and VQ-VAE models are trained using MSE loss to minimize reconstruction error. In addition, VQ-VAE applies a VQ loss to reduce the discrepancy between the encoder's output latent vector \(\vb*{z}_e\) and its nearest vector from a predefined codebook. This encourages the model to use discrete latent codes, ensuring consistency during reconstruction. The total VQ loss is given by:
\begin{align}
\mathcal{L}_{\mathrm{VQLoss}} &= \mathcal{L}_{\mathrm{vq}} + \mathcal{L}_{\mathrm{commit}} \nonumber \\
&= \left\|\operatorname{sg}[\vb*{z}_e(\vb*{H}_s)] - \vb*{e}\right\|_2^2 
+ \beta \left\|\vb*{z}_e(\vb*{H}_s) - \operatorname{sg}[\vb*{e}]\right\|_2^2,
\end{align}
where \(\vb*{z}_e(\vb*{H}_s)\) is the encoder output, \(\vb*{e}\) is the nearest codebook vector \(z_q\), and \(\beta\) is a hyperparameter controlling the weight of the commitment loss. The stop-gradient operation \(\operatorname{sg}[\cdot]\) prevents gradients from flowing through the enclosed variable, treating it as a constant during optimization.

The final loss function for VQ-VAE combines the MSE loss and the VQ loss:
\begin{equation}
\mathcal{L}_{\mathrm{VQVAE}} = \mathcal{L}_{\mathrm{MSE}} + \mathcal{L}_{\mathrm{VQLoss}}.
\end{equation}

\section{Experiments} \label{sec:exp}

\subsection{Simulation Setup}
\begin{table}[!h]   
\centering
\resizebox{.85\columnwidth}{!}{\begin{minipage}[t]{.9\columnwidth}
\centering
\begin{tabular} {l l}
\toprule[1.5pt]
\textbf{\textit{Parameter}} & \textbf{\textit{Value (Type)}} \\
\cmidrule(lr){1-1} \cmidrule(lr){2-2}
\# of TX$\times$RX antennas & 4$\times$16\\
Antenna config. & ULA with $0.5$ Ant. spacing \\
Carrier frequency & 40 [GHz] \\
\# of frame, Symbol per subframe & 1, 14 \\ 
\# of OFDM symbols & 140 \\
Subcarrier spacing & 15 [kHz] \\
Bandwidth & 10 [MHz] \\
\# of Subcarriers & 624 \\
Channel (train) & CDL-C \\
Channel (test) & CDL-$\{$A$\sim$D$\}$ \\
Delay spread, Max. Doppler shift & 30 [ns], 30 [Hz]\\
\bottomrule[1.5pt]
\end{tabular}

\end{minipage}}
\caption{Parameters for dataset.}
\label{table_paramter_dataset}
\end{table} 

    
    

\begin{figure*}[!th]
\centering
\begin{tabular}{c c c c}
    \multirow{2}{*}[-.5em]{%
        \begin{subfigure}[]{.23\linewidth}
        \centering
        \includegraphics[width=\linewidth]{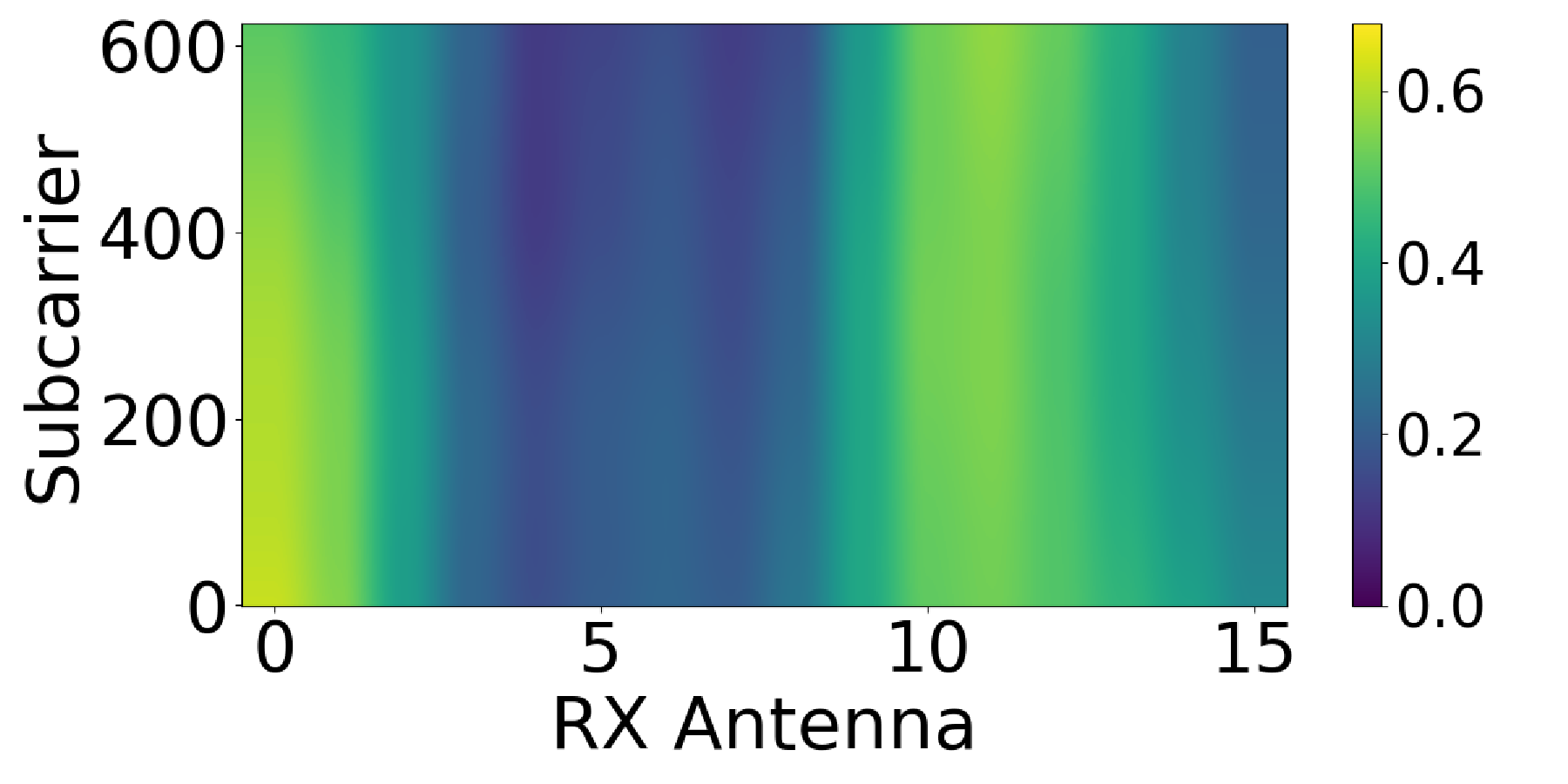}
        \caption{Ground Truth}
        \end{subfigure}} 
    & \begin{subfigure}[]{.23\linewidth}
    \centering
    \includegraphics[width=\linewidth]{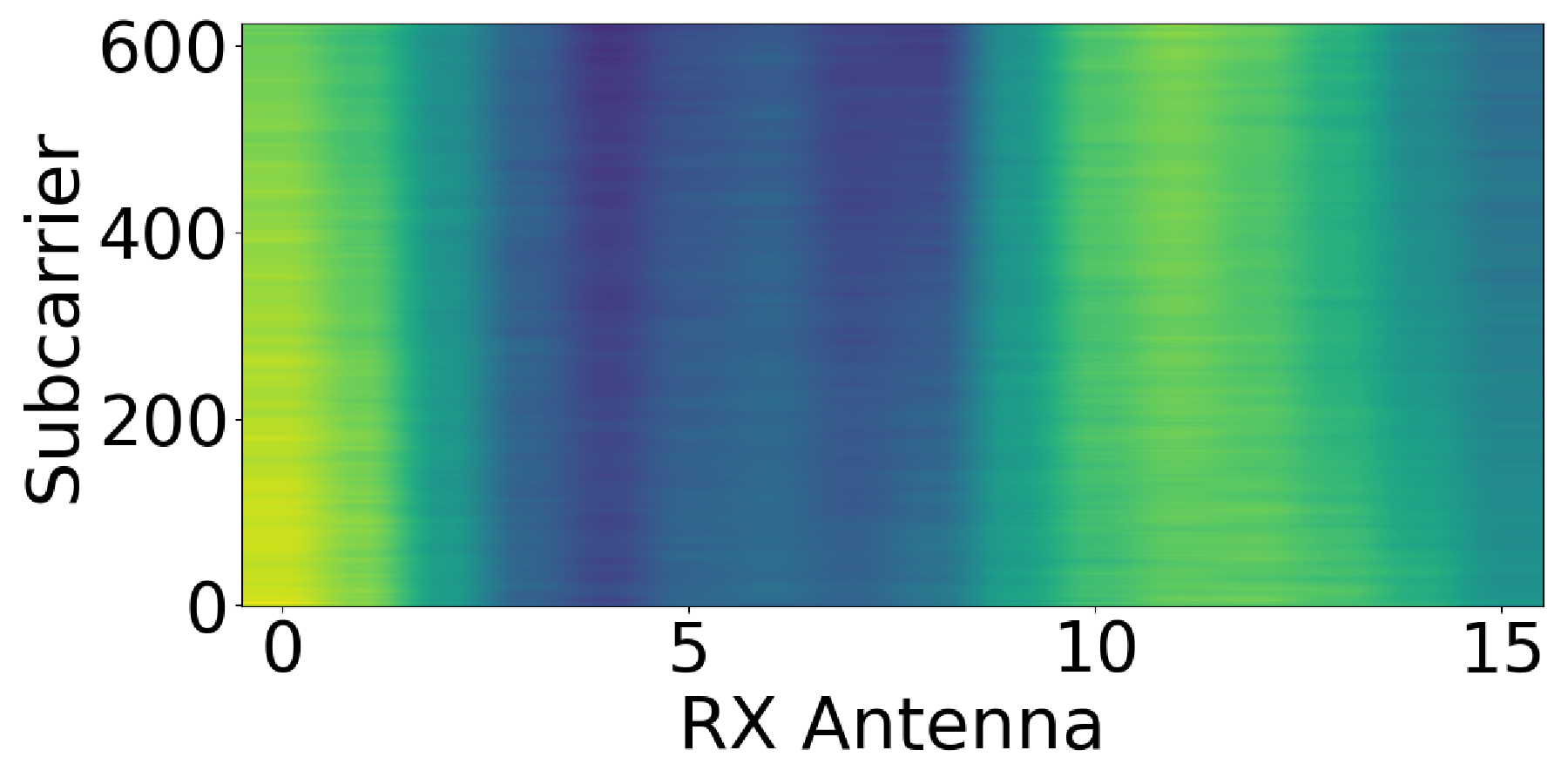}
    \caption{AE ($\gamma=30$ [dB])}
    \end{subfigure} 
    & \begin{subfigure}[]{.23\linewidth}
    \centering
    \includegraphics[width=\linewidth]{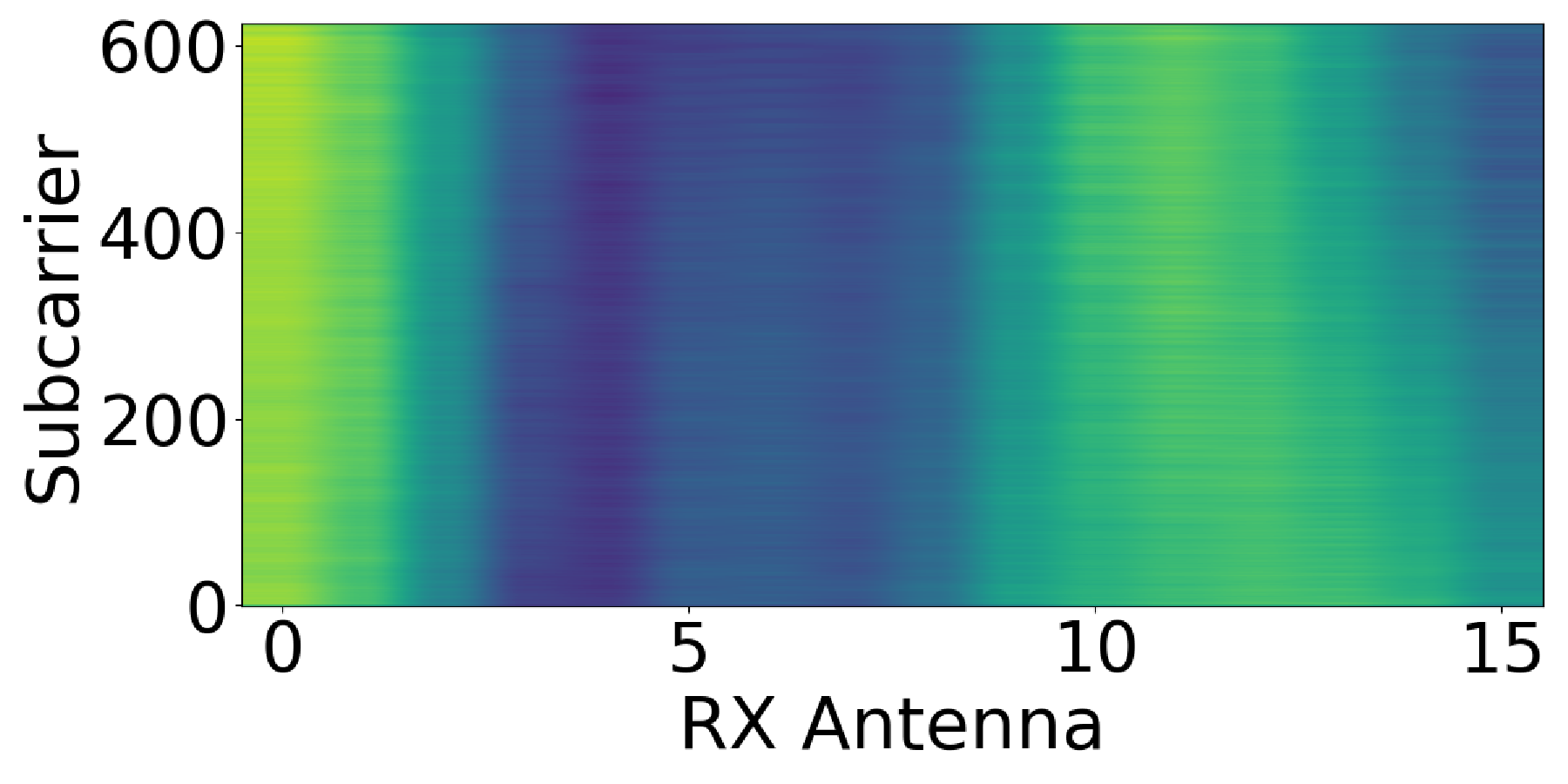}
    \caption{VAE ($\gamma=30$ [dB])}
    \end{subfigure} 
    & \begin{subfigure}[]{.23\linewidth}
    \centering
    \includegraphics[width=\linewidth]{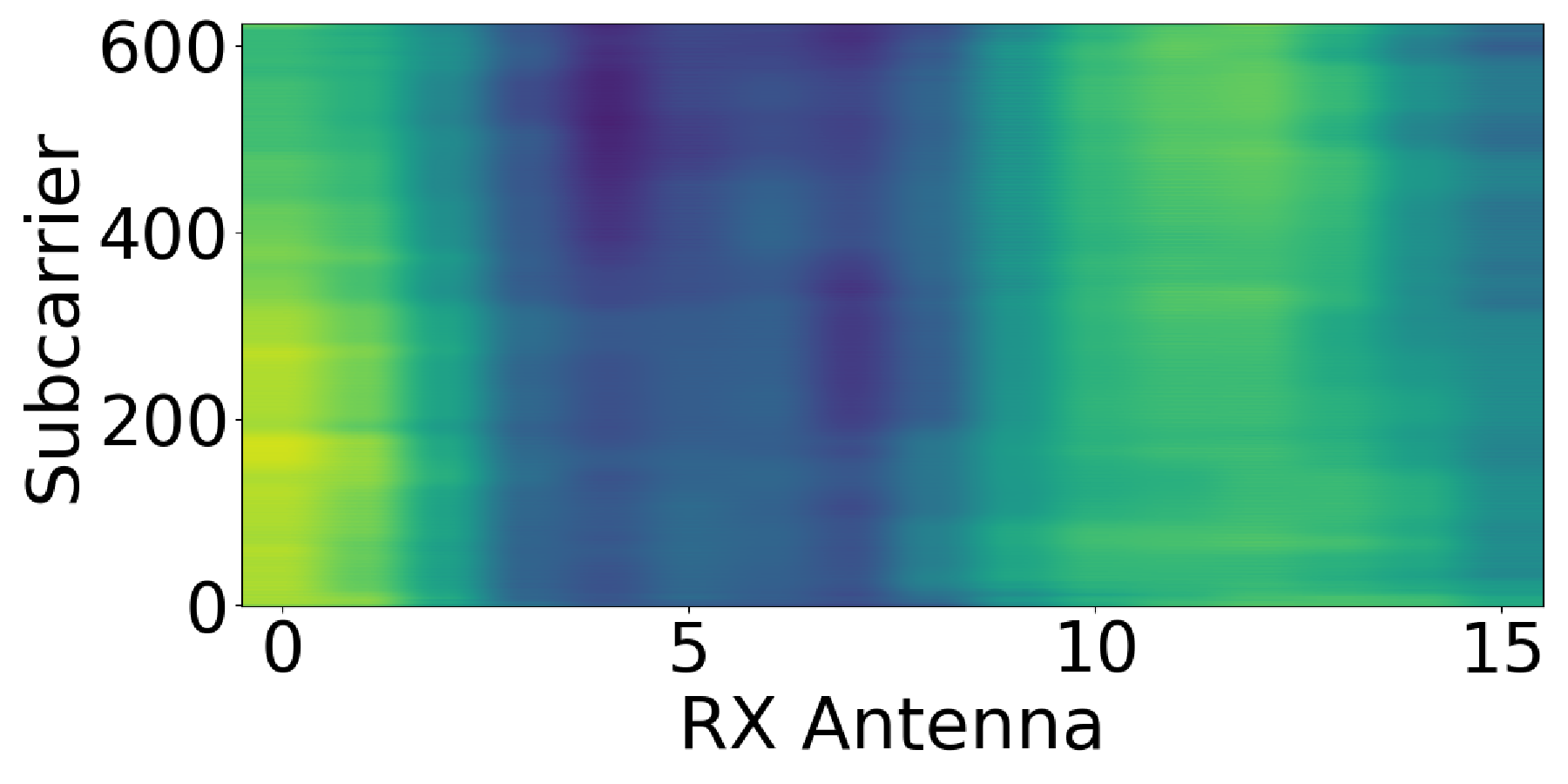}
    \caption{VQ-VAE ($\gamma=30$ [dB])}
    \end{subfigure} \vspace{1.em} \\     
    & \begin{subfigure}[]{.23\linewidth}
    \centering
    \includegraphics[width=\linewidth]{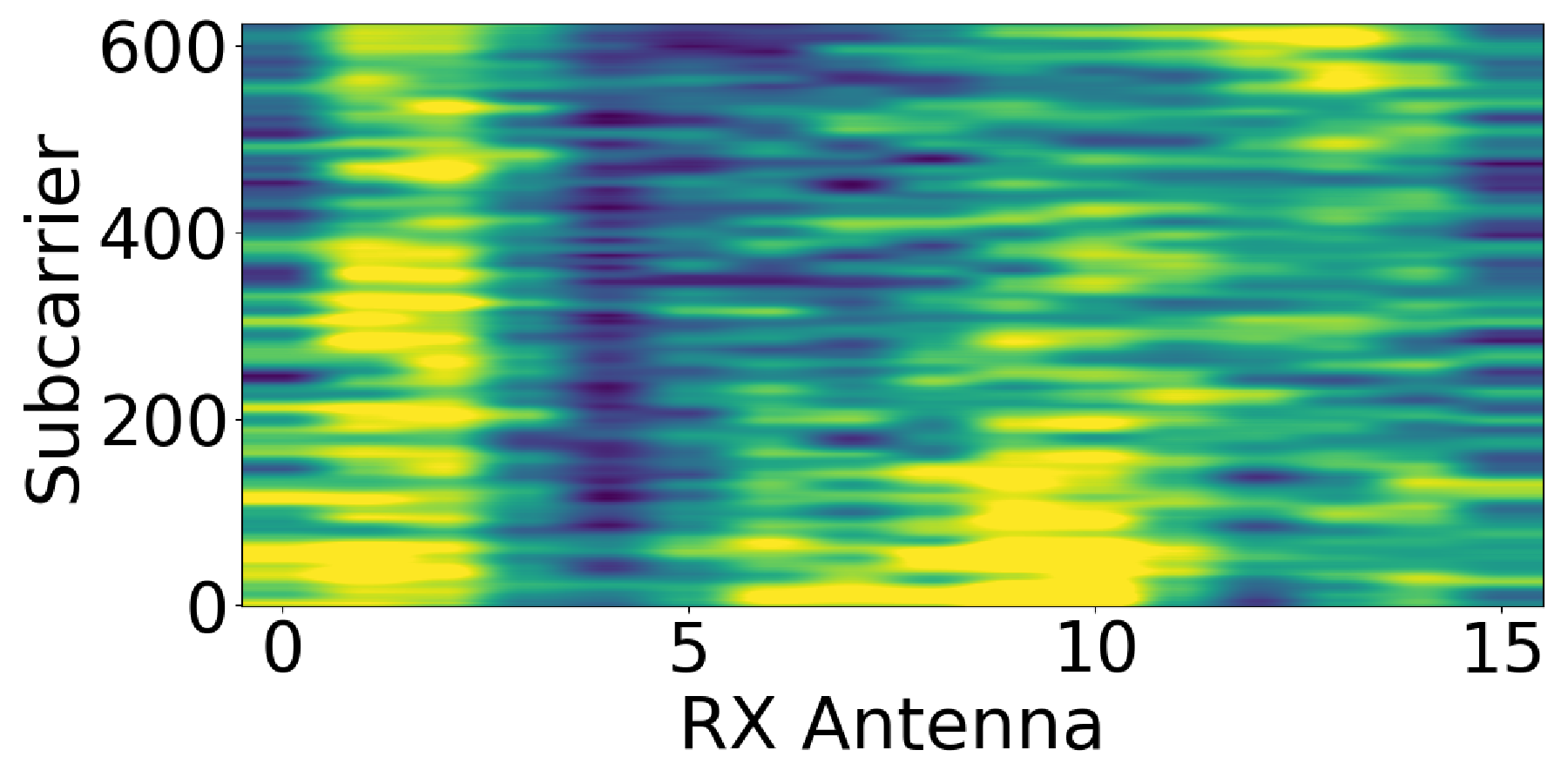}
    \caption{AE ($\gamma=0$ [dB])}
    \end{subfigure} 
    & \begin{subfigure}[]{.23\linewidth}
    \centering
    \includegraphics[width=\linewidth]{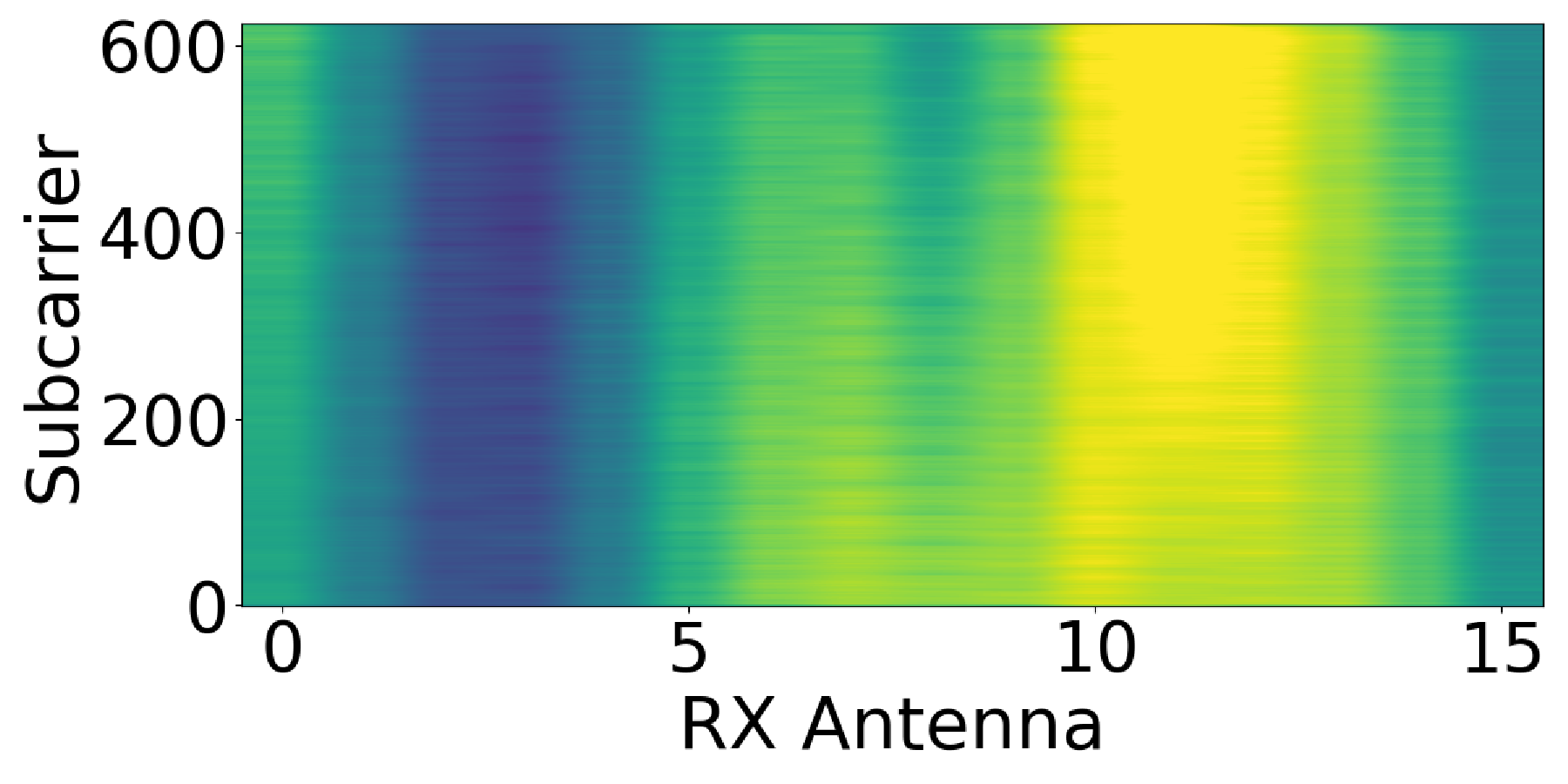}
    \caption{VAE ($\gamma=0$ [dB])}
    \end{subfigure} 
    & \begin{subfigure}[]{.23\linewidth}
    \centering
    \includegraphics[width=\linewidth]{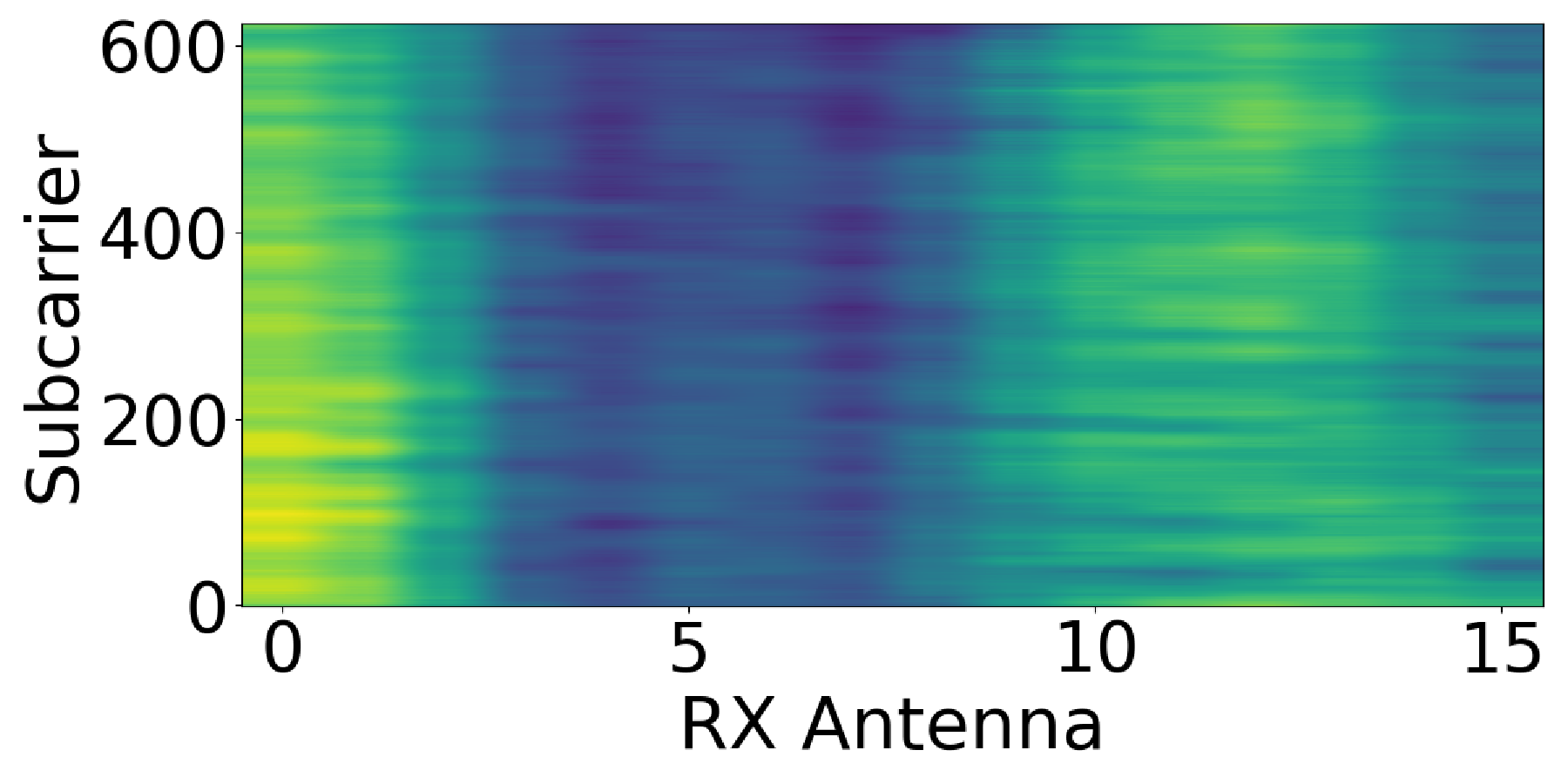}
    \caption{VQ-VAE ($\gamma=0$ [dB])}
    \end{subfigure} \\
\end{tabular}

\caption{Prediction results from AE, VAE, and VQ-VAE under good ($\gamma=30$ [dB]) and noisy ($\gamma=0$ [dB]) channel conditions, compared with the ground truth (Latent dim. = 64).}
\label{fig:prediction}
\end{figure*}

\BfPara{Dataset: 3GPP CDL-Family Channels} 
For training and evaluation, we utilize 3GPP CDL-family channel models (TR 38.901) \cite{TR38901}. All models are trained on 9000 samples from the CDL-C model and evaluated using 1000 samples from the CDL-$\{$A$\sim$D$\}$ channels. Table~\ref{table_paramter_dataset} provides a detailed summary of the channel and transceiver configurations.

\BfPara{Baselines} 
We compare VQ-VAE with the following baseline models:  
\begin{enumerate}
    \item \textbf{AE} \cite{ae}: A predictive model composed of convolutional (\texttt{Conv}) layers, which compress the input into a latent representation for data reconstruction, such as predicting missing antenna channels.
    \item \textbf{VAE} \cite{vae}: A generative model with a structure similar to AE, but introducing latent space regularization by parameterizing the space with mean and log-variance layers, enabling sampling and data generation.
    \item \textbf{Diffusion (DDPM) \cite{ddpm}}: A generative model using a UNet architecture to iteratively add and remove Gaussian noise from input data, achieving high-quality data generation through progressive denoising.
    \item \textbf{VQ-VAE (Proposed)} \cite{vqvae}: A generative model with an encoder, decoder, and vector quantizer, replacing latent vectors with the closest codebook entries to enhance reconstruction via quantized representation.
\end{enumerate}
The parameter configurations for each model are summarized in Table~\ref{table_paramter_baseline}.
\begin{table}[!h]   
\resizebox{.8\columnwidth}{!}{\begin{minipage}[t]{.9\columnwidth}
\centering
\begin{tabular} {l c c c}
\toprule[1.5pt]
\textbf{\textit{Parameter}} & \multicolumn{3}{c}{\textbf{AE, VAE, VQ-VAE}} \\
\cmidrule(lr){1-1} \cmidrule(lr){2-4}
Embedding dimension & \multicolumn{3}{c}{64}\\
Optimizer & \multicolumn{3}{c}{Adam}\\
Learning rate & \multicolumn{3}{c}{0.001 - 0.0005}\\
Batch size & \multicolumn{3}{c}{32, 64}\\
Epochs & \multicolumn{3}{c}{100} \\
\midrule
\textbf{\textit{Parameter}} & \textbf{AE} & \textbf{VAE} & \textbf{VQ-VAE} \\
\cmidrule(lr){1-1} \cmidrule(lr){2-2} \cmidrule(lr){3-3} \cmidrule(lr){4-4}
\# of embedding & N/A & N/A & 512 \\
KL weight & N/A & 0.000025 & N/A \\
Loss function & MSE & MSE + KL Div. & MSE + VQ Loss \\
\bottomrule[1.5pt]
\end{tabular}

\end{minipage}}
\caption{Parameters for AE, VAE, and VQ-VAE.}
\label{table_paramter_baseline}
\end{table} 

\begin{figure}[!h]
    \centering
    \includegraphics[width=.7\linewidth]{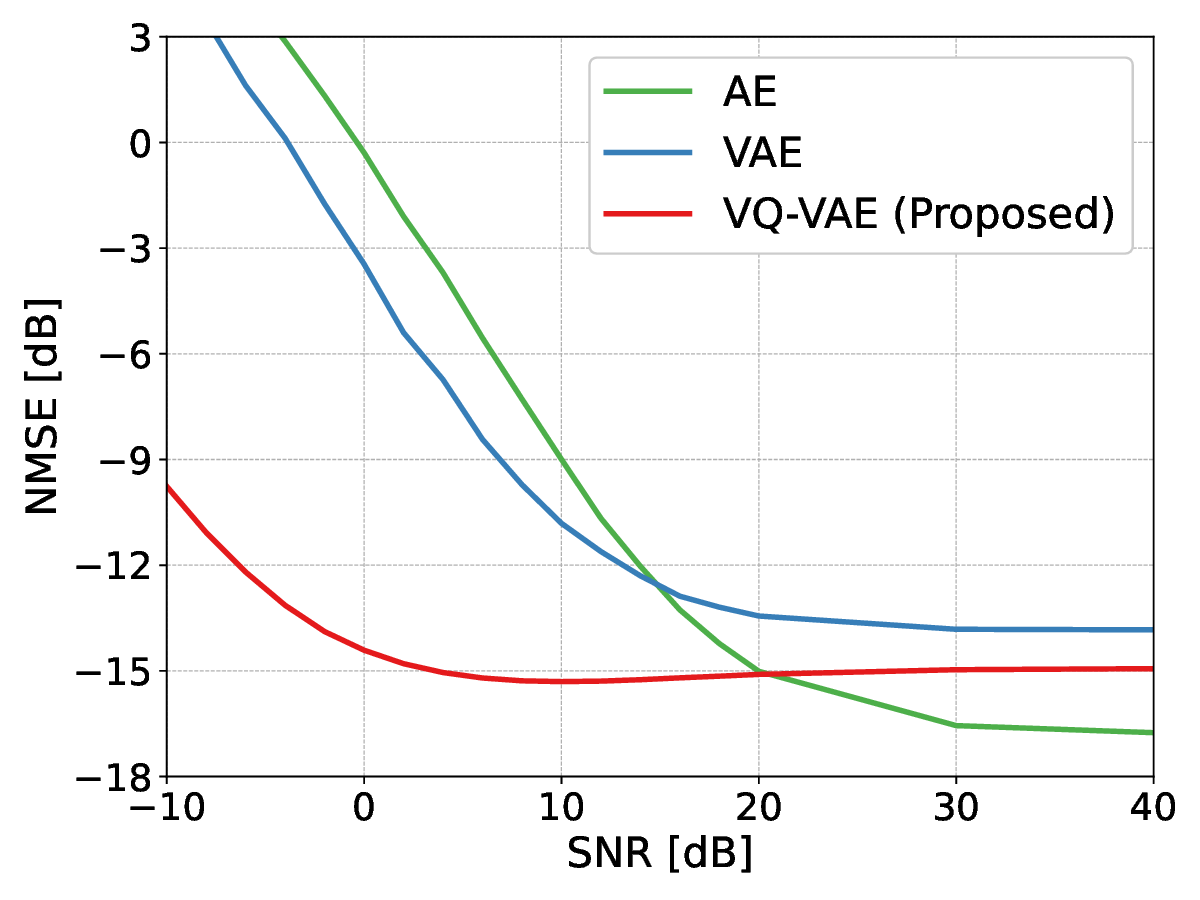}
    \caption{Comparative Analysis: NMSE vs. SNR [dB] for AE, VAE, and VQ-VAE.}
    \label{fig:comparison}
\end{figure}

\subsection{Simulation Results} \label{sec:Num}

\BfPara{Comparison} 
Figure~\ref{fig:comparison} shows the NMSE [dB] performance of AE, VAE, and the proposed VQ-VAE across different SNR levels. At high SNRs, all models achieve similar NMSE values around $-15$ [dB]. In contrast, under low-SNR conditions (e.g., $\gamma < 5$ [dB]), VQ-VAE demonstrates superior performance, maintaining an NMSE near $-10$ [dB], highlighting its robustness against noise.

\BfPara{Prediction}
Figure~\ref{fig:prediction} presents prediction samples from the baseline models. As observed in Figure~\ref{fig:comparison}, under favorable channel conditions (e.g., $\gamma = 30$ [dB]), all models provide accurate predictions. However, under noisy conditions (e.g., $\gamma = 0$ [dB]), only the VQ-VAE model provides reliable predictions, consistent with previous results.

\begin{figure*}[!h]
\centering
    \begin{subfigure}[]{.23\linewidth}
    \centering
    \includegraphics[width=\linewidth]{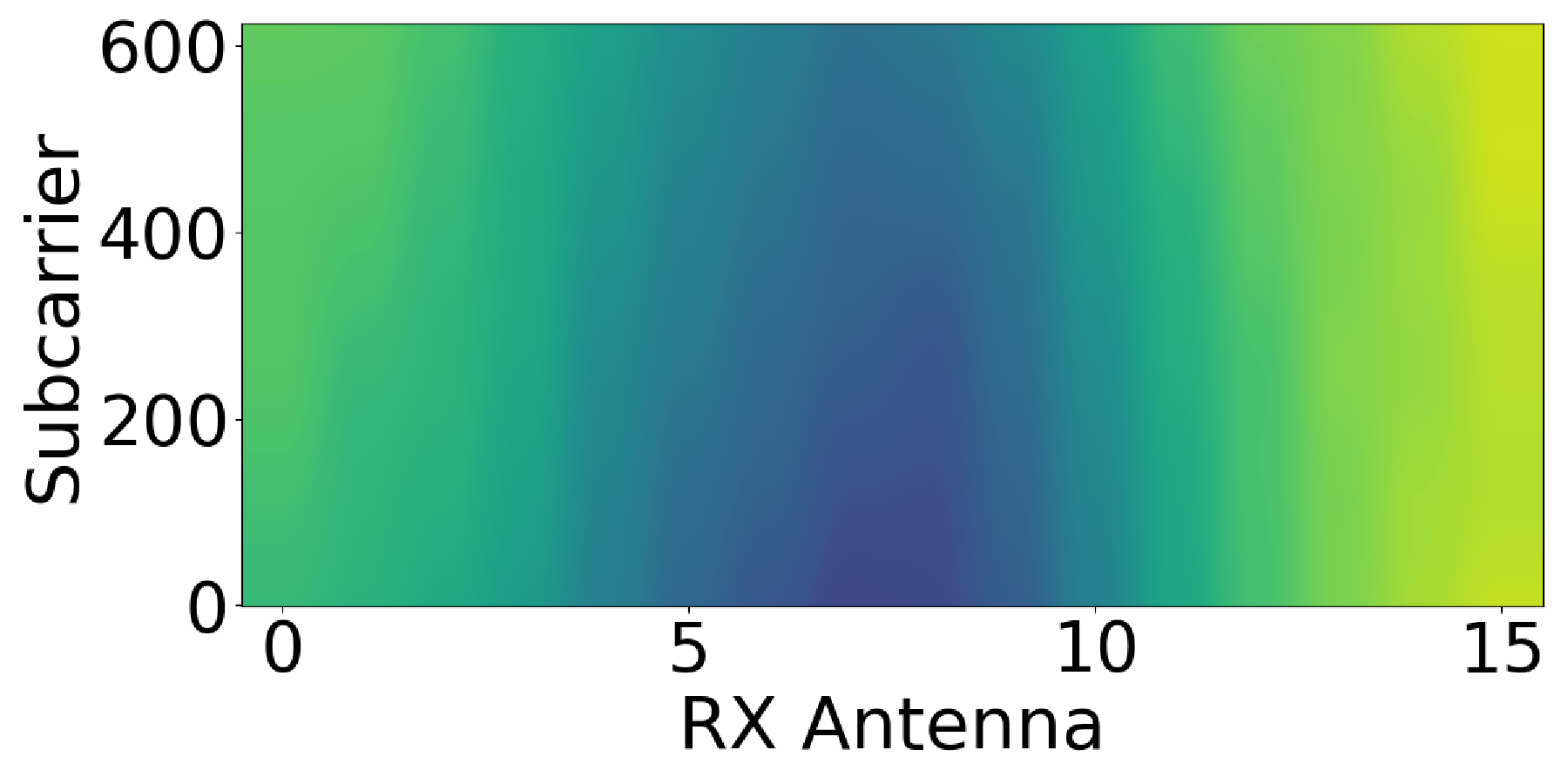}
    \caption{Ground-Truth (CDL-A)}
    \end{subfigure}
    \begin{subfigure}[]{.23\linewidth}
    \centering
    \includegraphics[width=\linewidth]{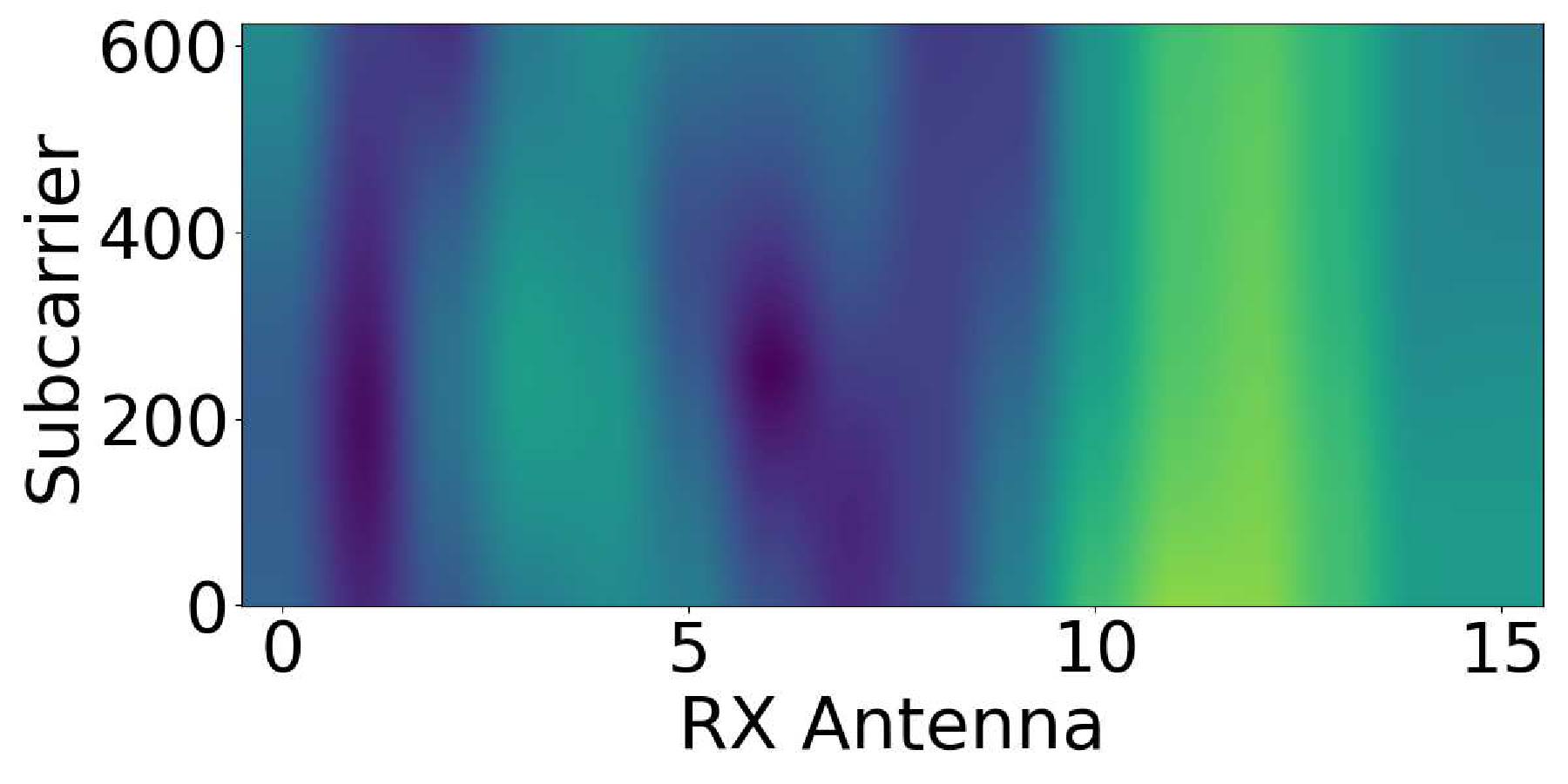}
    \caption{Ground-Truth (CDL-B)}
    \end{subfigure}
    \begin{subfigure}[]{.23\linewidth}
    \centering
    \includegraphics[width=\linewidth]{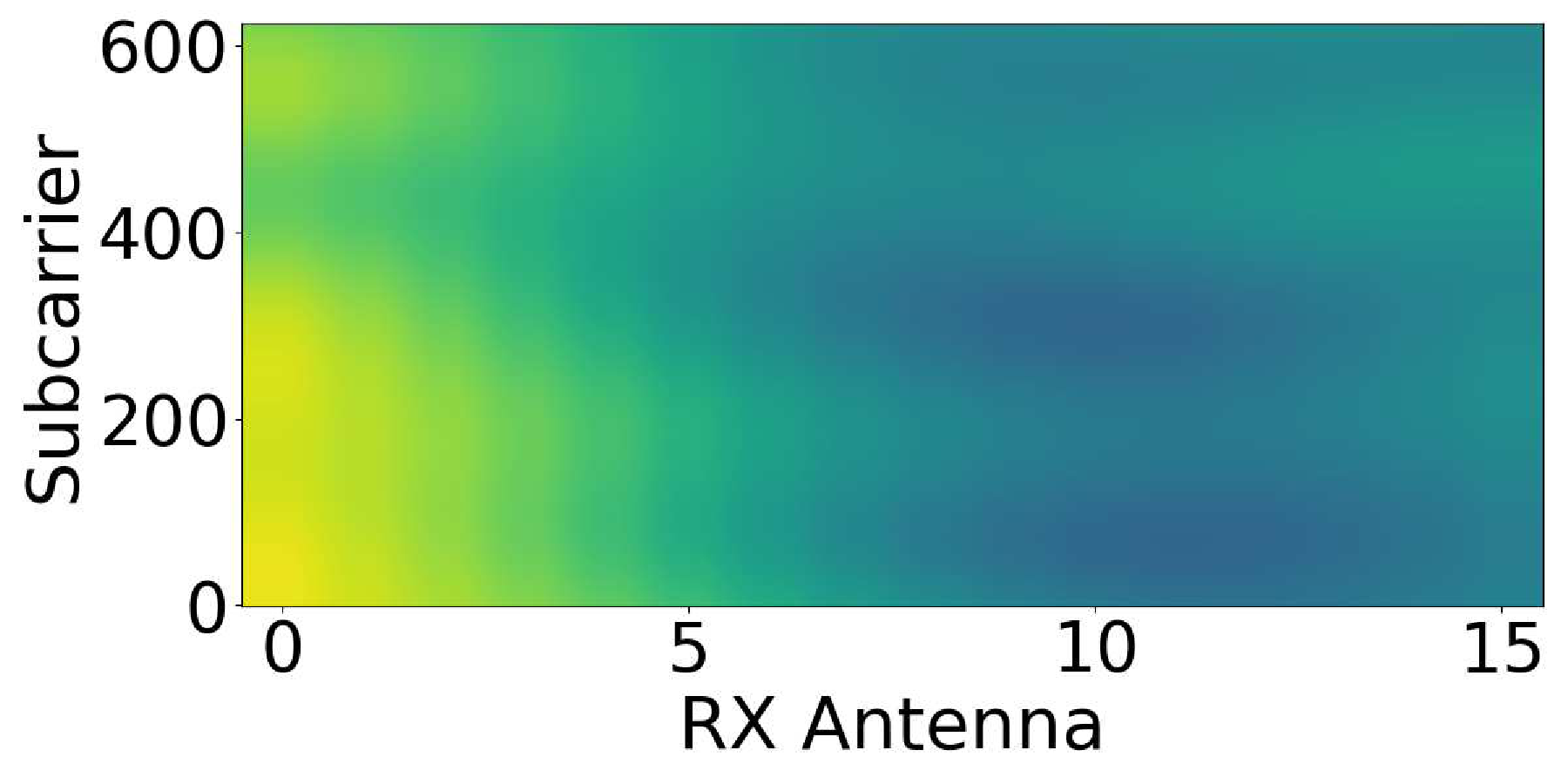}
    \caption{Ground-Truth (CDL-D)}
    \end{subfigure}

    \vspace{1.em}
    \begin{subfigure}[]{.23\linewidth}
    \centering
    \includegraphics[width=\linewidth]{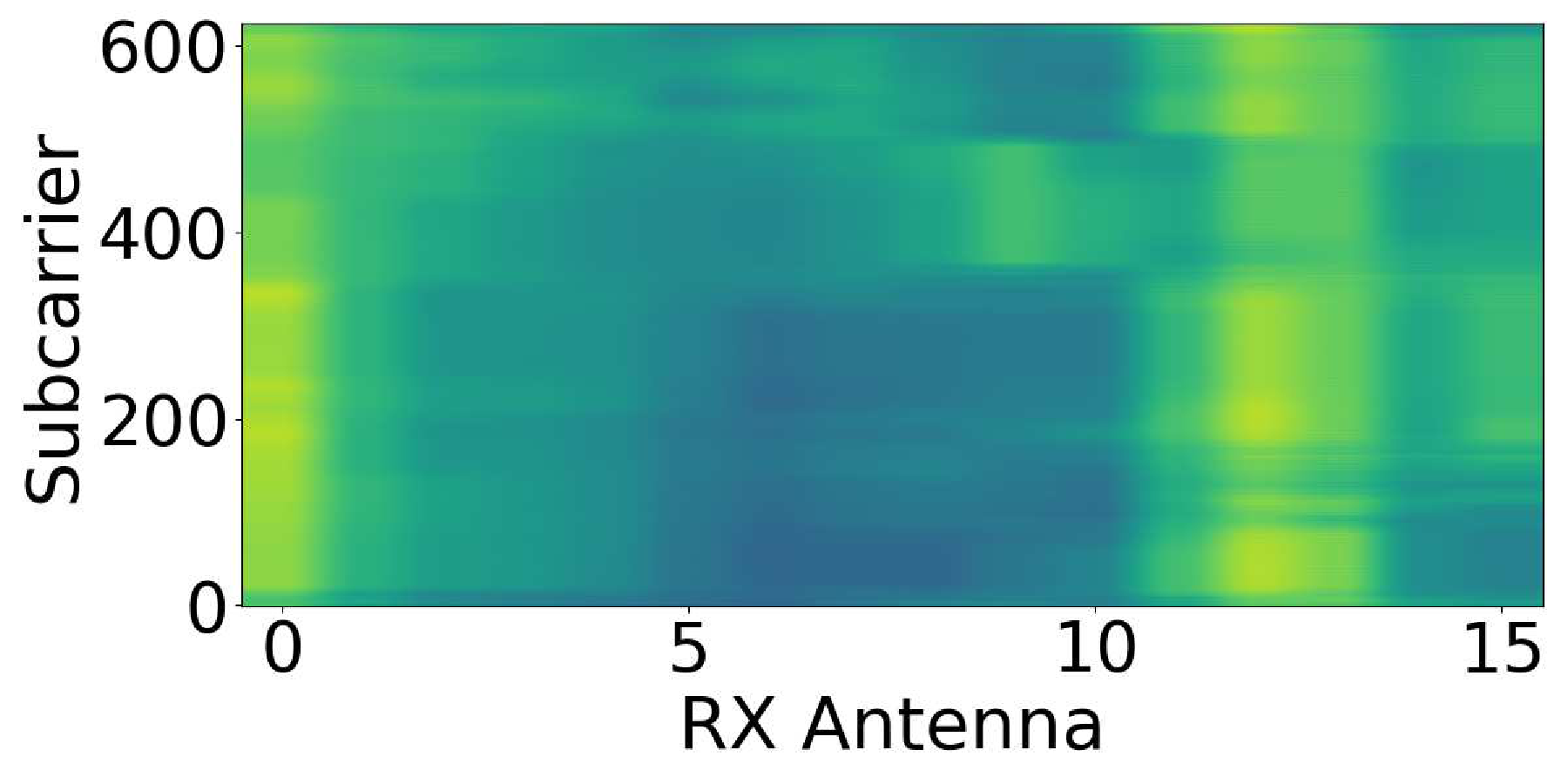}
    \caption{Prediction (CDL-A)}
    \end{subfigure}
    \begin{subfigure}[]{.23\linewidth}
    \centering
    \includegraphics[width=\linewidth]{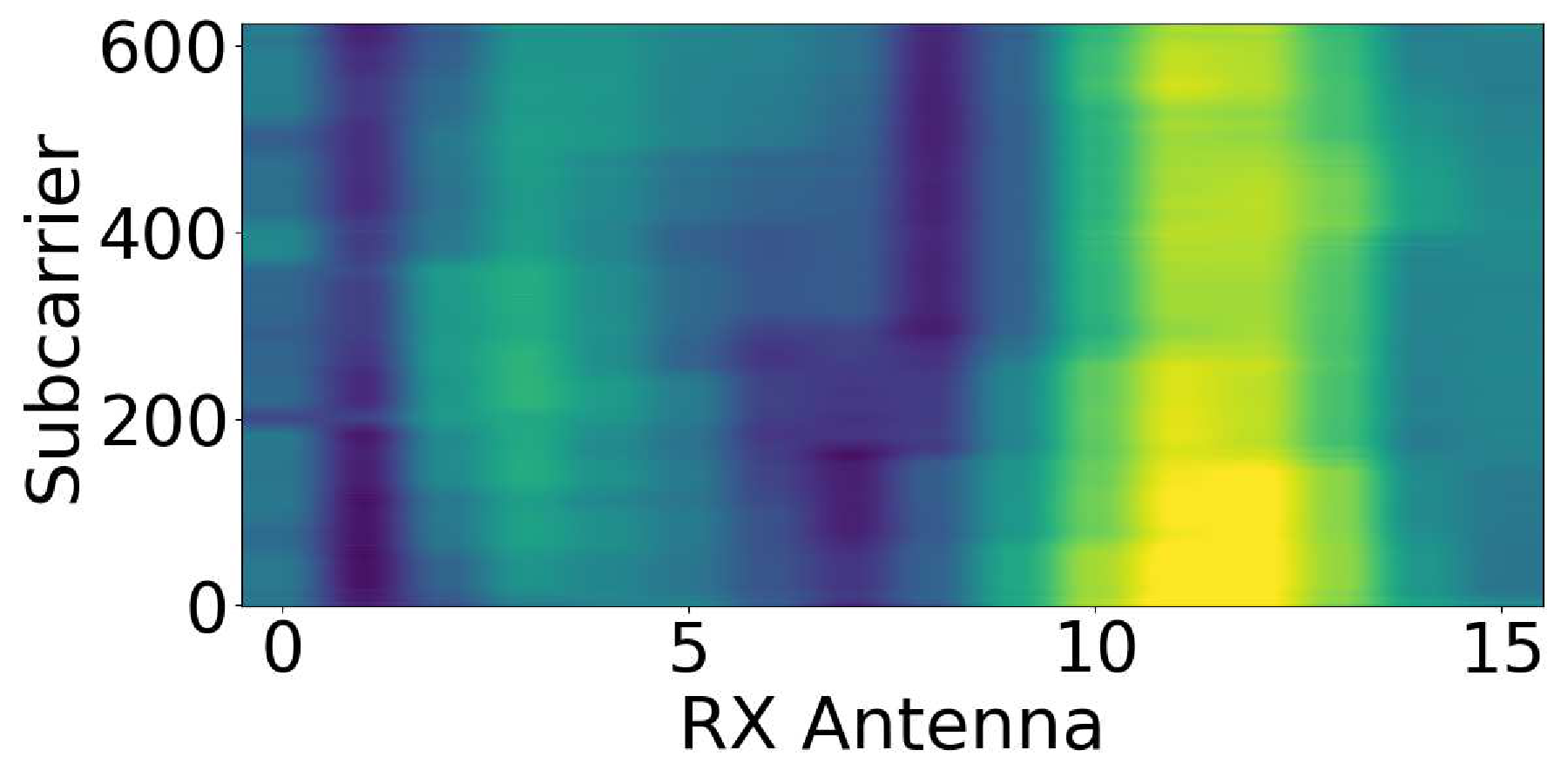}
    \caption{Prediction (CDL-B)}
    \end{subfigure}
    \begin{subfigure}[]{.23\linewidth}
    \centering
    \includegraphics[width=\linewidth]{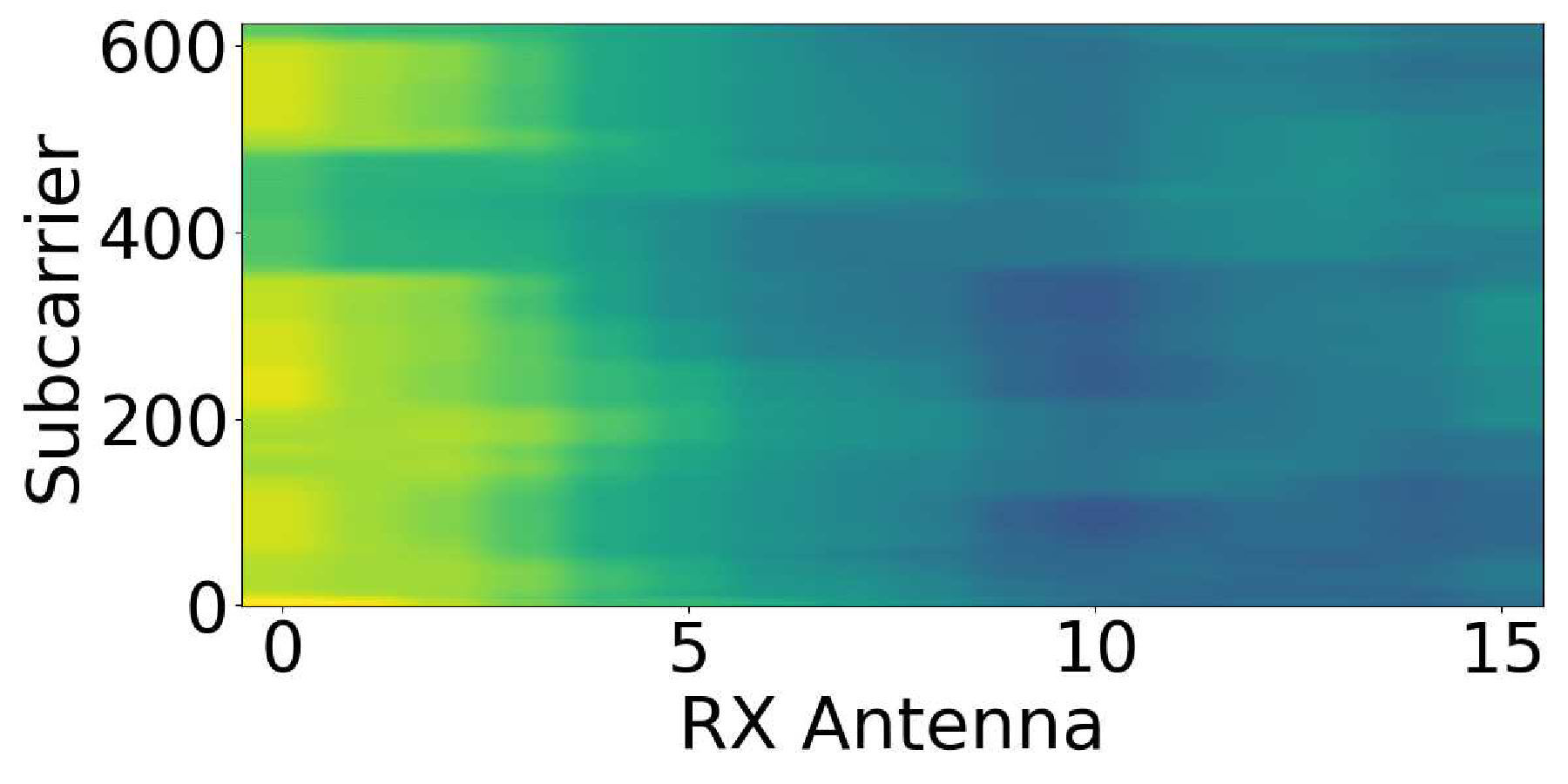}
    \caption{Prediction (CDL-D)}
    \end{subfigure}
    
\caption{Prediction results of the proposed VQ-VAE on Out-Of-Distribution (OOD) channels ($\gamma=30$ [dB], Latent dim. = 64). The top row shows the ground-truth, while the bottom row presents the corresponding predictions.}
\label{fig:prediction_ood}
\end{figure*}

\BfPara{Generalization Capability}
Fig. \ref{fig:prediction_ood} demonstrates the generalization capabilities of the proposed VQ-VAE using three Out-of-Distribution (OOD) datasets: CDL-A, CDL-B, and CDL-D. The results underscore VQ-VAE's ability to effectively predict on unseen channel data, although there appears to be potential for further improvement.
    
    
\begin{table*}[!ht]
\centering          
\resizebox{1.3\columnwidth}{!}{\begin{minipage}[h]{1.25\columnwidth}
\centering
\begin{tabular}{c | c | c | c | c}
    \toprule[1.5pt]
    & \textbf{\textit{Inference time}} [ms] & 
    \textbf{\textit{Memory}} [MB] &
    \textbf{\textit{Training time}} [s] & \textbf{\textit{Memory}} [MB] \\
    \midrule
    \textbf{AE} & 3.374 & 62.67 & 514 & 216 \\
    \textbf{VAE} & 4.442 & 108.59 & 840 & 469 \\
    \textbf{VQ-VAE} & 5.340 & 175.20 & 1576 & 1099 \\
    \cmidrule(lr){1-5} 
    \textbf{Diffusion (DDPM)} & 122.624 & 1385.41 & 30491 & 36135 \\
    \bottomrule[1.5pt]
\end{tabular}

\end{minipage}}
\centering
\caption{Inference time, training time, and memory usage for AE, VAE, and VQ-VAE.}
\label{table:complexity}
\end{table*}
\BfPara{Complexity}
Table~\ref{table:complexity} compares the computational complexity of VQ-VAE with baseline models (AE, VAE, and DDPM) in terms of inference time, training time, and memory usage. All experiments were conducted on an NVIDIA A100 GPU, with inference measured using a single sample and training time evaluated over 9000 training and 1000 validation samples.

VQ-VAE incurs higher computational costs than AE and VAE due to its vector quantization step but remains significantly more efficient than the diffusion model (DDPM). AE is the most lightweight, while VAE introduces moderate overhead from latent space regularization. DDPM’s iterative generation imposes heavy time and memory demands, making it less suitable for real-world deployment.

While VQ-VAE achieves strong performance in noisy conditions with manageable overhead, further optimizations are required for seamless integration into practical systems.

\section{Conclusions} \label{sec:conclusion}
This work introduced a VQ-VAE-based generative model for robust mMIMO cross-antenna channel prediction. The proposed model demonstrated superior performance, particularly in noisy conditions, achieving up to 15 [dB] NMSE gains over standard AEs and 9 [dB] over VAEs. These results highlights the effectiveness of generative models in addressing uncertainties in noisy wireless communication systems.

Future research could explore the application of generative models for tasks such as channel estimation and feedback compression in mMIMO systems. Further efforts may focus on optimizing model architectures to enhance the trade-off between prediction accuracy and computational efficiency.

\vspace{1.em}
{\bf Acknowledgements:} Part of this work was funded by NSF under grants 2133655 and 2008443. 

\bibliographystyle{IEEEbib}
\bibliography{refs}
\end{document}